\documentclass[prb,twocolumn,a4paper,floatfix]{revtex4-1}

\usepackage{latexsym}
\usepackage{graphicx}
\newcommand{\figwidth}{2.7 in}

\newcommand{\figwidthb}{3.8 in}
\usepackage{subfigure}    
\usepackage{verbatim}
\usepackage{amsmath}
\usepackage{color}
\usepackage{xcolor}
\usepackage{soul}

\begin{document}

\title{Effects of ``stuffing'' on the atomic and electronic structure of the pyrochlore Yb$_2$Ti$_2$O$_7$} 
\author{Soham S. Ghosh$^{(1,2)}$}
\author{ Efstratios Manousakis$^{(1,2,4)}$}
\affiliation{
$^{(1)}$ Department  of  Physics, 
  Florida  State  University,  Tallahassee,  Florida  32306-4350,  USA\\
$^{(2)}$ National High Magnetic Field Laboratory,
  Florida  State  University,  Tallahassee,  Florida  32306-4350,  USA\\
$^{(4)}$Department   of    Physics,   University    of   Athens,
  Panepistimioupolis, Zografos, 157 84 Athens, Greece
}
\date{\today}
\begin{abstract}
 There are reasons to believe that the ground state of the magnetic rare earth pyrochlore Yb$_2$Ti$_2$O$_7$ is 
 on the boundary between competing ground states. We have carried out \textit{ab initio} density functional 
 calculations to determine the most stable chemical formula as a function of the oxygen chemical potential and 
 the likely location of the oxygen atoms in the unit cell of the ``stuffed" system.  We find that it is energetically 
 favorable in the ``stuffed" crystal (with an Yb replacement on a Ti site) to contain oxygen vacancies which 
 dope the Yb 4\textit{f} orbitals and qualitatively change the electronic properties of the system. 
 In addition, with the inclusion of the contribution of spin-orbit-coupling (SOC) on top of the GGA+U 
 approach, we investigated the electronic structure and the magnetic moments of the most stable ``stuffed" system. 
  In our determined ``stuffed" structure  the valence bands as compared to those of the pure system are pushed down 
  and a change in hybridization between the O 2\textit{p} orbitals and the metal ion states is found. 
  Our first-principle findings should form a foundation for effective models describing the low-temperature 
  properties of this material whose true ground state remains controversial.
\end{abstract}

\pacs{}
\maketitle

\section{Introduction}
 
Pyrochlore materials with the general formula A$_2$B$_2$O$_7$, where $\mathrm{A}$ is a trivalent rare earth  
ion and $\mathrm{B}$ is a transition-metal atom, display a diverse set of physical properties\cite{Gardner2010}. 
In the pyrochlore structure, the A and B cations form two distinct interpenetrating lattices of corner-sharing 
tetrahedra. In such a geometry, the natural tendency to form long-range magnetic order is frustrated and 
hence these materials have been the subject of considerable theoretical and experimental interest for the 
last two decades\cite{Gardner2010}. 
Some pyrochlores such as Ho$_2$Ti$_2$O$_7$ and Dy$_2$Ti$_2$O$_7$ with large local Ising-like magnetic 
moments are believed to be long range dipolar\cite{Hertog2000} ``spin-ice" compounds.\cite{Bramwell2001,Clancy2009} 

In comparison to these ``classical" pyrochlores, the properties of 
Yb$_2$Ti$_2$O$_7$ are more debatable. An extensive body of 
experimental\cite{Yasui2003,Chang2012,Chang2014,Lhotel2014,Gaudet2016,Hodges2002,Yaouanc2003,Gardner2004,Ross2009,D'Ortenzio2013,Bhattacharjee2016} 
and theoretical\cite{Ross2011,Applegate2012,Hayre2013} research have claimed 
 Yb$_2$Ti$_2$O$_7$ states of different nature. It is known to have a ferromagnetic Curie-Weiss temperature 
of $\sim 0.65$ K and a first-order transition at $\sim 0.24$ K\cite{Hodges2002}.
Its properties are dominated by the 4$f^{13}$ electrons of the Yb$^{+3}$ ions which form the A 
sublattice tetrahedra network, as shown in Fig.~\ref{fig:tetrahedra1}(a). Large spin-orbit coupling (SOC)
and crystal field create a lowest Kramer's doublet state energetically separated 
from the first excited doublet by $620$ K\cite{Hodges2001}. The low-energy spin-dynamics 
as revealed by inelastic neutron scattering\cite{Ross2011} can be modelled by 
an effective pseudospin-1/2 moment\cite{Onoda2011}. The magnetic interaction 
of Yb$_2$Ti$_2$O$_7$ has been described in the $S = 1/2$ subspace by an anisotropic 
exchange Hamiltonian\cite{Ross2011} with four independent exchange constants\cite{Curnoe2007}.

Ross~et.~al\cite{Ross2012} have pointed out the role of off-stoichiometry in the 
difference in ground state properties of this material in various studies, which we 
must take into account. Generally speaking, it was found that 
systems with a single specific heat anomaly between 214 mK and 265 mK\cite{Ross2011,Yaouanc2011,blote1969}
are stoichiometric, whereas systems with broad humps in specific heat\cite{Ross2011,Yaouanc2011,Chang2012} 
are best described as  ``stuffed", with Yb substitution on some Ti sites. Stuffing introduces some new 
nearest neighbor interactions among the Yb$^{+3}$ ions as well as different oxygen environments, 
which is expected to  significantly influence the 
ground state properties.

While density functional theory (DFT) has been used to study  
 this system taking the active $f$ electrons into account\cite{Deilynazar2015}, 
there is a general need for ab-initio calculations to study the effect of off-stoichiometry in this class of 
materials and to form a basic microscopic understanding of the ground-state properties, which we shall 
provide here. It is well-known that materials with partially filled $f$ orbitals 
 pose serious challenges to DFT calculations, especially of their magnetic properties. 
 Furthermore, the complexity and range of the various possible interactions, as well as the very small energy 
scales of the exchange Hamiltonian\cite{Ross2011} put a detailed picture of the magnetic 
order in the ground state beyond the scope of first-principles calculations. However, one should still 
 start from the atomic and electronic structure suggested by DFT calculations. In particular, 
 an ab-initio investigation of the most stable atomic structure in the case of ``stuffing'' can be very useful information.
In addition, the electronic structure and the description of oxygen-vacancy related change in valence 
states would be pertinent information in building a model for this pyrochlore system. 

In this paper, we employ DFT computations to investigate the 
atomic, electronic and magnetic properties of stoichiometric and stuffed Yb$_2$Ti$_2$O$_7$. 
We consider stuffing at various concentrations and find that all the stuffed systems considered are unstable towards 
formation of oxygen vacancies for a wide range of oxygen chemical potential. The stable vacancy created 
in the neighborhood of the substituted cation site leads to significant changes in the structural and 
electronic properties as compared to the stoichiometric system. We identify Ti$^{+3}$ oxidation states as a 
consequence of the stuffing and note that the Ti atoms do not possess significant magnetic moments. We apply the 
Hubbard-based DFT+U approach to account for the Coulomb interactions between electrons and find that a minimum value of U 
in Yb \textit{f} orbitals is required to account for the correct electronic phase of Yb$_2$Ti$_2$O$_7$. 

The paper is organized as follows: After a brief explanation 
of our computational methods in Sec.~\ref{sec:method}, we present our 
detailed results in Sec.~\ref{sec:results} and present our conclusions in Sec.~\ref{sec:conclusion}.

\section{Computational Methods}
\label{sec:method}
 
 \begin{figure}[htp]
\vskip 0.2 in
\begin{center}
\includegraphics[width=\figwidthb]{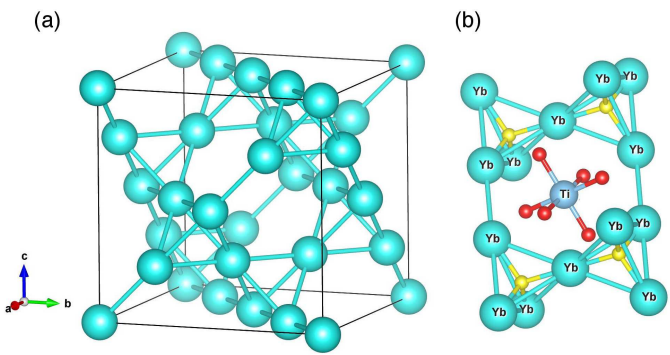}
\end{center}
\caption{ 
         (a) The pyrochlore crystal structure of Yb$_2$Ti$_2$O$_7$ showing the corner sharing 
         tetrahedral Yb network in a conventional simple cubic lattice (Ti and O atoms have been omitted for clarity).
         (b) One Yb hexagon with a Ti atom at its center forming part of a $\left \{ 111 \right \}$ Kagome plane. 
         This Ti atom is replaced by an Yb atom in the stuffed system. Also shown are some selected 
         Yb tetrahedra connected to the Kagome plane and perpendicular to it. One of 
         the oxygen atoms the Ti is bonded to (shown in red) is removed to create an Wyckoff \textit{f}-site oxygen vacancy. The 
         oxygen atom (shown in yellow) at the center of the Yb tetrahedra is removed to create an Wyckoff \textit{b}-site 
         vacancy.}
\label{fig:tetrahedra1}
\vskip 0.2 in
\end{figure}
 
We perform Spin-GGA and Spin-GGA+U computations using plane-wave basis set (cutoff of 520 eV) 
with the projected augmented wave methodology\cite{Blochl} used to describe the wavefunctions of the 
electrons as implemented in the \textsc{VASP} package\cite{Shishkin3,Fuchs,Shishkin2,Shishkin1}. 
We use the Perdew-Burke-Ernzerhof (PBE) formulation of the exchange correlation functional\cite{PBE}.
 The 4\textit{s}, 3\textit{d}, 3\textit{p} electrons of titanium 
and the  2\textit{s}, 2\textit{p} electrons of the oxygen
were treated as valence electrons. For ytterbium, the 4\textit{f}, 6\textit{s} as well 
as the semi-core 5\textit{s}, 5\textit{p}, 4\textit{s} electrons were included.
Yb$_2$Ti$_2$O$_7$ belongs to the space group Fd$\bar{3}$m (227) with Yb and Ti both forming a lattice of corner sharing 
tetrahedra. The Yb tetrahedral network is shown in Fig.~\ref{fig:tetrahedra1}(a). There are two 
types of oxygen in the unit cell, one in the Wyckoff \textit{f}-site, and the 
other type in the Wyckoff \textit{b}-site at the center of the Yb tetrahedra.  
We consider simulation cells with 2, 4 and 8 Yb$_2$Ti$_2$O$_7$ formula units (f.u.) to study the role of 
stuffing and oxygen vacancy stability at various concentrations.
The stoichiometric system consists of four ytterbium atoms in a 22-atom primitive unit cell. 
For the Yb $f$-orbitals in these systems, we vary the parameter U between 0 and 6 eV to see its effect on the electronic structure,
and use values of 0, 0.8 and 1 eV for the on-site exchange J. We use U $= 2$ eV 
and J $= 0.8$ eV for the Ti 3$d$ orbitals. Changing the values of U and J for Ti 
does not have a significant effect on the system properties.
For all other systems, we fix the values of U and J 
at U$_{\mathrm{Yb}} = 6$ eV, J$_{\mathrm{Yb}} = 1$ eV, U$_{\mathrm{Ti}} = 2$ eV and J$_{\mathrm{Ti}} = 0.80$ eV. 

 All the stuffed systems are structurally relaxed until the forces 
were converged to less than 10 meV/\AA\, for each ion.
The stuffed systems have additional Yb atoms substituting 
Ti atoms at the center of the hexagon in the $\left \{ 111 \right \}$ Kagome plane (Fig.~\ref{fig:tetrahedra1}(b)). 
We create vacancies in the stuffed systems by removing an oxygen atom either in the Wyckoff \textit{f}-site (neighbor of this Ti atom) 
or in the Wyckoff \textit{b}-site (center of the Yb tetrahedron) to study their stability and their effect 
on the electronic structure. 

The Brillouin zones for the stoichiometric systems with 2 Yb$_2$Ti$_2$O$_7$ f.u. in the computational cell, and 
its stuffed variants (with and without an oxygen vacancy) were each sampled with a $ 7 \times 7 \times 7$ \textit{k}-point 
grid. By increasing the k-point sampling up to $ 9 \times 9 \times 9$ for selected systems, 
we found the energy converged to within 0.20 meV/atom. The Brillouin zone of the 4 Yb$_2$Ti$_2$O$_7$ f.u. systems  
were sampled with up to  $ 6 \times 6 \times 3$ \textit{k}-point grids 
and that of the 8 Yb$_2$Ti$_2$O$_7$ f.u. systems were sampled with $ 3 \times 3 \times 3$ \textit{k}-point grids.

In the 2 and 4 f.u. systems, we included SOC to accurately represent the electronic structure.
Symmetry was turned off altogether when SOC was included, except during the ionic relaxation cycles which were 
carried out without SOC taken into consideration. SOC in these systems generate the Dzyaloshinsky-Moriya (DM) 
interaction between nearest neighbor Yb ions. Even though SOC in 4\textit{f} orbitals is strong, 
weak \textit{f}-orbital hopping makes the DM interaction relatively small, as can be seen from the value of 
the $J_{4}$ parameter in the anisotropic exchange Hamiltonian needed in Ref.~\onlinecite{Ross2011} in order to fit the neutron scattering data. 
From this point of view, structural relaxation without including SOC is justifiable.
SOC was not included in computing the oxygen vacancy stability in the 8 Yb$_2$Ti$_2$O$_7$ f.u. systems.

\section{Atomic, electronic and magnetic structure} 
\label{sec:results}
\subsection{Pure pyrochlore Yb$_2$Ti$_2$O$_7$ crystal}

\begin{figure}[htp]
\vskip 0.2 in
\begin{center}
\includegraphics[width=\figwidth]{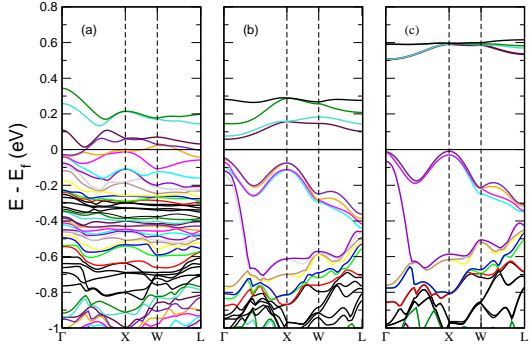}
\end{center}
\caption{
         Electronic band structure of Yb$_2$Ti$_2$O$_7$ calculated using Spin-GGA + U + SOC method,
         for various values of U and J in Yb \textit{f} orbitals,
         shown along the crystal high symmetry lines. We use in (a) $U = 0$ eV, $J = 0$ eV , which gives 
         us a metallic ground state. In (b), we use $U = 3$ eV, $J = 0.7$ eV, which opens a small gap. In 
         (c), we use $U = 6$ eV, $J = 1$ eV. 
} 
\label{fig:4Yb_bands}
\vskip 0.2 in
\end{figure}

For 4\textit{f} electrons of the Yb$^{+3}$ ions, SOC is strong and should be considered before 
crystal field effects. The thirteen \textit{f} electrons should have spin $S = 1/2$ and angular momentum $L = 3$ according to 
Hund's rule. The fourteen-fold degeneracy is lifted by SOC leading to eight-fold degenerate $J = 7/2$ 
states and six-fold degenerate $J = 5/2$ states. The crystal field further splits the $J = 7/2$ states into 
four Kramer's doublet and the ground state can be described by an effective $S = 1/2$ pseudospin. The thirteen 
electrons fill this manifold creating one empty conduction band. Thus, we expect as many 4\textit{f}-character bands 
near the Fermi level as there are Yb atoms in the computational unit cell. The stuffed systems will be expected to have more 
conduction bands than the stoichiometric system.

In Fig.~\ref{fig:4Yb_bands}(a), we plot the bands for stoichiometric Yb$_2$Ti$_2$O$_7$ with four Yb atoms in the 
unit cell.
We find that for low correlations in the 4\textit{f} orbitals, 
the ground state is metallic. This differs from the experimental result that Yb$_2$Ti$_2$O$_7$ 
is an insulator which is either dark red in color or clear transparent\cite{Arpino2017}. 
A moderate value of $U \sim 3$ eV is needed to open up a gap in the electronic spectrum, which we show 
in Fig.~\ref{fig:4Yb_bands}(b). Once the system becomes an insulator, the qualitative features do not 
change with the value of U. For large correlation ($U = 6$ eV, $J = 1$ eV), we find in 
Fig.~\ref{fig:4Yb_bands}(c) a direct band gap of $0.52$ eV. This is close to that in a previous 
study\cite{Deilynazar2015} with the same $U$ and $J$ values where the band gap was found to be $0.6$ eV. 
The ground state is that a single 4\textit{f} hole at each Yb site, occupying one of the conduction bands 
in Fig.~\ref{fig:4Yb_bands}(b or c) and the O 2\textit{p} occupied sites shown as the valence states in 
the same figures. The anisotropic superexchange interaction between Yb ions is driven by a virtual 
hooping\cite{Onoda2011} between these levels.

It may be noted that a previous DFT work\cite{Xiao2007}, in which the \textit{f} 
electrons were frozen in the core during the calculations, 
found the difference between the conduction band minimum (CBM) and valence band maximum (VBM) to be $3.34$ eV. 
In our work, that corresponds to the difference of $3.25$ eV between the valence band maximum and the 
empty Ti 3\textit{d} states (shown later in Fig.~\ref{fig:totaldos}).

\subsection{The stuffed Yb$_2$Ti$_2$O$_7$ with oxygen vacancy}

To study the stability of oxygen vacancies in stuffed Yb$_2$Ti$_2$O$_7$ We compare the energies of  
Yb$_{2+x}$Ti$_{2-x}$O$_{7-\delta}$ with Yb stuffing concentration 
of $25\, \%$ ($x = 1/2$) and $12.5\, \%$ ($x = 1/4$) with the corresponding ``$\delta = 0$" states. 
We expect our description to be quantitatively accurate for high stuffing concentrations or for 
microscopic clusters of stuffing in the crystal, and qualitatively meaningful in the dilute 
regime. To create these systems we replace Ti atoms at sites  
labeled in Fig.~\ref{fig:tetrahedra1}(b) by Yb atoms. Due to the substitution of a ${+4}$ Ti ion with 
a ${+3}$ Yb ion, it is reasonable to expect one oxygen atom bonded to the cation in this site to leave. 
While there are direct evidences of Yb stuffing\cite{Ross2012,Arpino2017}, 
oxygen vacancy, which seems quite plausible given general oxidation state arguments, has 
been either assumed or argued for indirectly\cite{Mostaed2017} based on a Ti valence change. 
To find the stability of creation of oxygen vacancy, we compute:
\begin{align}
F = \Bigl [ F_{V} - F_{0} + \mathrm{\mu_{O}} \Bigr ] /N_u, 
\end{align}
where $F_{V}$ is the free-energy of the stuffed unit cell with an oxygen vacancy, $F_0$ is the free-energy of the 
stuffed no-vacancy unit cell, $\mu_{\mathrm{O}}$ is the oxygen chemical potential, and $N_u$ is the number 
of Yb octahedra included in our computational unit cells.

We use the following computational cells in our stability computations. $25\, \%$ Yb stuffing 
and one oxygen vacancy in the 22-atom (2 f.u.) unit cell 
creates a system with the chemical formula Yb$_{2+x}$Ti$_{2-x}$O$_{7-x}$ ($x = 1/2$). To create a unit cell with the chemical 
formula Yb$_{2+x}$Ti$_{2-x}$O$_{7-x/2}$,  we replace two Ti atoms with a Yb atom each, and remove one oxygen atom. 
This creates a stuffing concentration of $25\, \%$ ($x = 1/2$) in the 44-atom (4 f.u.) unit cell 
and $12.5\, \%$ ($x = 1/4$) in the 88-atom (8 f.u.) unit cell.
To find the possible range of $\mathrm{\mu_{O}}$, we first note that
the value is bounded from above by the chemical potential of formation of the triplet O$_2$ molecule,
\begin{align*}
\mathrm{\mu_{O}} < \mathrm{\mu_{O_2}/2} = -4.917\, \mathrm{eV}. 
\end{align*}
To find the lower limit, we assume that $\mathrm{\mu_{O}}$ is the same in the gas phase and in the stuffed pyrochlore.
Then the chemical potentials of the elements in the stuffed pyrochlore are constrained by:
\begin{align*}
(2+x)\mathrm{\mu_{Yb}} + (2-x)\mathrm{\mu_{Ti}} + 7\mathrm{\mu_{O}}
= \mathrm{\mu_{Yb_{2+x}Ti_{2-x}O_{7}}}.
\end{align*}
To stop phase separation into Yb$_{bulk}$ and Ti$_{bulk}$, the chemical potentials of 
Yb and Ti themselves must satisfy inequality constraints:
\begin{align*}
\mathrm{\mu_{Yb}} < \mathrm{\mu_{Yb}^{bulk}} = -1.231\, \mathrm{eV}, \\
\mathrm{\mu_{Ti}} < \mathrm{\mu_{Ti}^{bulk}} = -7.898\, \mathrm{eV}. 
\end{align*}
Using the above inequality constraints and the computed value $\mathrm{\mu_{Yb_{(2+x)}Ti_{(2-x)}O_{7}}} = -79.395\, \mathrm{eV}$, 
the allowable energy of the oxygen chemical potential is
\begin{align*}
-9.210\, \mathrm{eV} < \mathrm{\mu_{O}} < -4.917\, \mathrm{eV}. 
\end{align*}
\begin{figure}[htp]
\vskip 0.2 in
\begin{center}
\includegraphics[width=\figwidth]{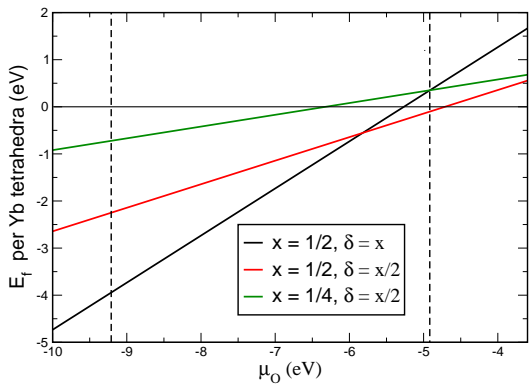}
\end{center}
\caption{
         Free energy of the oxygen vacancy process in stuffed Yb$_2$Ti$_2$O$_7$ for $25\, \%$ and $12.5\, \%$ Yb stuffing. 
         Three different types of unit cells with chemical formula Yb$_{2+x}$Ti$_{2-x}$O$_{7-\delta}$ are shown 
         and their free energies are compared with those of the corresponding stuffed structures with no oxygen vacancy. 
         The vacancies are at Wyckoff \textit{f}-sites. The limits of oxygen chemical potential is shown by dotted lines. We use $U_{Yb} = 6$ eV, $J_{Yb} = 1$ eV, $U_{Ti} = 2$ eV, $J_{Ti} = 0.8$ eV. 
} 
\label{fig:stability}
\vskip 0.2 in
\end{figure}
As shown in Fig.~\ref{fig:stability}, we find that the pyrochlore structures at all stuffing concentrations considered will 
create an oxygen vacancy over a wide range of possible oxygen chemical potential. We further find that the 
Wyckoff \textit{f}-site oxygen vacancy is more stable than the Wyckoff \textit{b}-site vacancy by $0.4$ eV per Yb tetrahedra 
so we do not consider the latter in further discussions. If Ti cations outside the pyrochlore phase are 
created by the synthesis process which produces off-stoichiometric ratios of Yb and Ti, the oxygen ions will 
react to produce a stable TiO$_2$ phase.

\begin{figure}[htp]
\vskip 0.2 in
\begin{center}
\includegraphics[width=\figwidth]{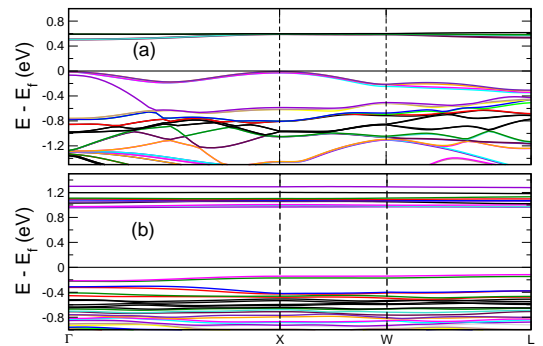}
\end{center}
\caption{
         (a) Electronic band structure of stoichiometric Yb$_2$Ti$_2$O$_7$ calculated using Spin-GGA + U + SOC 
         method (Fig~\ref{fig:4Yb_bands}(c) repeated).
         Electronic band structure of Yb$_{2+x}$Ti$_{2-x}$O$_{7-x/2}$ ($x = 1/2$), 
         shown along the crystal high symmetry lines. The vacancy is in the Wyckoff \textit{f}-site. 
         The ten energy bands around the Fermi level are 4\textit{f}
         in character. Results are for $U_{Yb} = 6$ eV, $J_{Yb} = 1$ eV, $U_{Ti} = 2$ eV, $J_{Ti} = 0.8$ eV. 
 }
\label{fig:Ovac_bands}
\vskip 0.2 in
\end{figure}

The replacement of Ti atoms with Yb and the removal of an oxygen atom cause the stuffed Yb to move away from
the vacant oxygen site, displacing it from the center of the Kagome hexagon. For $x = 1/2$, the various Yb-O bond lengths for this Yb atom 
are found to be between $2.14$ \AA\, and $2.32$ \AA, which is shorter than the Yb-O bond length of $2.51$ \AA\, at the 
Wyckoff \textit{f}-site but longer than the Ti-O bond length of $1.94$ \AA. In contrast to the 4\textit{f} orbitals of Yb$^{3+}$ 
where the ground state changes from metallic to insulating as the value of Hubbard $U$ crosses a threshold, 
the role of $U$ in Ti 3\textit{d} orbitals is found to be  minimal in determining the character near the Fermi level. 
We show the electronic band structure of the stuffed pyrochlore Yb$_{2+x}$Ti$_{2-x}$O$_{7-x/2}$ ($x = 1/2$), 
which we earlier found to be stable, in Fig.~\ref{fig:Ovac_bands}(b). We repeat Fig.~\ref{fig:4Yb_bands}(c) in Fig.~\ref{fig:Ovac_bands}(a) for comparison. There are ten Yb 4\textit{f} conduction bands in Fig.~\ref{fig:Ovac_bands}(b). 
Due to the oxygen vacancy caused by stuffing, the mostly O 2\textit{p} valence bands just below the Fermi level are missing. 
Effectively, the O 2\textit{p} energy bands are thus pushed down compared to those in the the stoichiometric system. The band gap compared 
to the stoichiometric system is widened, to $1.02$ eV.

\begin{figure}[htp]
\vskip 0.2 in
\begin{center}
\includegraphics[width=\figwidth]{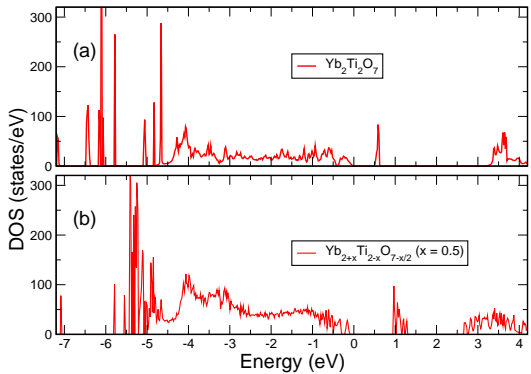}
\end{center}
\caption{
         (a) Electronic density of states of Yb$_2$Ti$_2$O$_7$ and (b) Yb$_{2+x}$Ti$_{2-x}$O$_{7-x/2}$ 
         at $x = 1/2$ calculated using Spin-GGA + U + SOC. The Fermi energies are set to zero. 
         Results are for $U_{Yb} = 6$ eV, $J_{Yb} = 1$ eV, $U_{Ti} = 2$ eV, $J_{Ti} = 0.8$ eV. 
} 
\label{fig:totaldos}
\vskip 0.2 in
\end{figure}

To show the full impact of off-stoichiometry, Fig.~\ref{fig:totaldos} compares the density of states (DOS) 
of the stoichiometric crystal with the stuffed ($x = 1/2$) crystal having an oxygen vacancy. The states near the Fermi 
surface are Yb 4\textit{f} whereas the valence states are mostly O 2\textit{p} states. 
As a consequence of removal of an oxygen atom, the Ti atom near the vacancy acquire a $+3$ valence 
character and some of the O 2\textit{p} valence states are absent. The increased hybridization between 
Ti 3\textit{d} states and O 2\textit{p} states show up in the stuffed crystal between $2.5$ and $4$ eV  (Fig.~\ref{fig:totaldos} (b)). 
Change in Yb-O bond length caused by the defects increases Yb-O hybridization which shows up as an increased density of 
states below the Fermi level.\\ 

With only the isotropic Heisenberg exchange (without SOC) the magnetic moments in the pyrochlore geometry are 
 frustrated. The DM interaction removes that frustration. However, classical DM interaction does not 
discriminate between the spin-ice (two-in, two-out) state and the all-in/all-out (all four moments pointing towards the 
center in one tetrahedron, and pointing out in the next tetrahedron). Furthermore, the coefficients in the exchange 
Hamiltonian evaluated by Ross and co-workers\cite{Ross2012} by fitting neutron scattering 
excitation data are exceedingly small, specially the $J_{4}$ constant ($ \sim 0.01$ meV) associated with the DM term. 
Our computation methods were accurate to within $0.2$ meV/atom. Predictably, we found all the states 
we considered as starting points (spin-ice, all-in/all-out, (100) ferromagnetic, (100) antiferromagnetic, 
(111) ferromagnetic and (111) antiferromagnetic) to be degenerate.

We find that the total magnetic moment per Yb atom in the stoichiometric system to 
be $0.86 \pm 0.005\, \mu$B and the orbital magnetic moment per Yb atom to be $0.25 \pm 0.007\, \mu$B. In the 
non-stoichiometric systems, we find the total magnetic moment per Yb atom to be $0.92 \pm 0.02\, \mu$B and orbital 
magnetic moment per Yb atom to be $0.24 \pm 0.02\, \mu$B.
These magnetic moments compare reasonably to experimental findings\cite{Hodges2002} and are different from a 
previous DFT calculation\cite{Deilynazar2015}. There was no noticeable magnetic moment in the Ti atoms even as 
there was a change in oxidation state due to the oxygen vacancy.

The magnetic moments were calculated using VASP non-self-consistent routines where the charge density was kept fixed 
during the non-collinear calculations. It is also possible to include non-collinear SOC in a fully self-consistent 
way. When we do that, the magnetic moments change significantly. In the stoichiometric system, we find total magnetic 
moment per Yb atom to be $0.28\, \mu$B and orbital magnetic moment to be $0.53\, \mu$B. This suggests a 
rotation of the orbital and spin expectation values with respect  to each other. 

\section{Conclusions}
\label{sec:conclusion}

The atomic and electronic structure of stoichiometric and 
stuffed Yb$_2$Ti$_2$O$_7$ are investigated using density functional calculations within the 
GGA+U scheme and including spin-orbit coupling. Through our stability computations, we show that once there is 
substitution of a Yb$^{+3}$ in a Ti$^{+4}$ site (the stuffed system), it is energetically favorable to 
contain oxygen vacancy defects in the Wyckoff \textit{f}-sites, leading to stable oxygen deficient materials 
characterized by a change in Ti valence state from Ti$^{+4}$ to Ti$^{+3}$. Stuffing and oxygen vacancy defects cause changes in 
the Ti-O and Yb-O bond lengths which create a different hybridization character. The O 2\textit{p} valence bands are pushed down   
 and the band gap is increased. We find the electronic phase and the band gap 
of Yb$_2$Ti$_2$O$_7$ to be dependent on the value of correlation U in the Yb \textit{f} orbitals.  
We notice no significant magnetic moment in the Ti atoms in the defect system. The magnetic moments of the Yb 
atoms were calculated and found to be reasonably close to measured values. Off-stoichiometry 
introduces a magnetic moment in the substituted lattice sites which have different exchange pathways as compared 
to the Yb tetrahedra in the stoichiometric system. This, combined with oxygen vacancy and structural distortion 
is likely to destroy long range magnetic order. While the GGA+U scheme seems to work reasonably well, it is 
known for its limitations in \textit{f}-electron systems. However, our revealed atomic structure of the ``stuffed" compound, 
in particular the stability of the oxygen vacancies, is expected to be correct since 
its determination involves quite large energy differences between it and the other competing structures.

Note added: When our lengthy computations were finished and we were preparing 
this manuscript for submission, we became aware of a recently published experimental 
paper, Ref.~\onlinecite{Mostaed2017}, which using 
scanning electron transmission microscopy (STEM) directly images the Yb substitution on Ti sites. 
It further argues the presence of oxygen vacancy by observing changes in Ti 
valence using electron energy loss near-edge structure (ELNES) spectra. These findings are consistent with 
our ab-initio findings and point to the role of disorder in determining the ground-state properties of Yb$_2$Ti$_2$O$_7$. 

\vskip 0.2 in
\section{Acknowledgments}
This work was supported in part by the U.S. National High Magnetic Field
Laboratory, which is partially funded by the NSF DMR-1157490 and 
the State of Florida.

\end{document}